# Time-Resolved Studies of Stick-Slip Friction in Sheared Granular Layers

S. Nasuno[1,2,*], A. Kudrolli[1,#], A Bak[1], and J.P. Gollub[1,3,+]

[1] Physics Department, Haverford College, Haverford PA 19041 and

[2] Department of Electrical Engineering, Kyushu Institute of Technology, Tobata, Kitakyushu 804-8550, Japan.

[3] Department of Physics, University of Pennsylvania, Philadelphia PA 19104, USA.

April 21, 1998

## Abstract

Sensitive and fast force measurements are performed on sheared granular layers undergoing stick-slip motion, along with simultaneous optical imaging. A full study has been done for spherical glass particles with a +-20% size distribution. Stick-slip motion due to repetitive fluidization of the granular layer occurs for low driving velocities. Between major slip events, slight creep occurs that is highly variable from one event to the next. The effects of varying the stiffness $k$ of the driving system and the driving velocity $V$ are studied in detail. The stick-slip motion is almost periodic for spherical particles over a wide range of parameters, whereas it becomes irregular when k is large and V is relatively small. At larger $V$, the motion becomes smoother and is affected by the inertia of the upper plate bounding the layer. Measurements of the period $T$ and amplitude $A$ of the relative motion are presented as a function of $V$. At a critical value $V_c$, a transition to continuous sliding motion occurs. The transition is discontinuous for $k$ not too large, and large fluctuations occur in the neighborhood of the transition. The time dependence of the instantaneous velocity of the upper plate and the frictional force produced by the granular layer are determined within individual slipping events. The frictional force is found to be a multi-valued function of the instantaneous velocity during slip, with pronounced hysteresis and a sudden drop just prior to resticking. Measurements of vertical displacement reveal a very small dilation of the material (about one tenth of the mean particle size in a layer 20 particles deep) associated with each slip event; the dilation reaches its maximum amplitude close to the time of maximum acceleration. Finally, optical imaging reveals that localized microscopic rearrangements precede (and follow) each macroscopic slip event; their number is highly variable, and the accumulation of these local displacements is associated with macroscopic creep. The behavior of smooth particles is contrasted qualitatively with that of rough particles.

# I. Introduction and Background

The dynamical properties of granular materials have been widely explored, but their complex behavior continues to challenge our understanding [1, 2]. The responses of these materials to applied forces, both compressional and shear, are important physical properties. This paper is concerned with the behavior of granular layers that are subjected to applied shear forces, and to the resulting frictional forces that are generated in the material. The response to imposed shear forces can be quite varied: the material can remain at rest, or it may yield in a way that leads to steady, fluctuating, or stick-slip motion. The material may become inhomogeneous, with parts behaving as a fluid while other parts remain solid-like. In fact one of the fascinating aspects of granular friction is that it involves both the solid and fluid aspects of the material.

Non-lubricated friction between solid surfaces, with or without an intervening layer of granular material, has been widely studied. Here, the frictional forces and the resulting dynamics are controlled by the asperities (protrusions) of the contacting surfaces, or the geometrical irregularity of the grains. These phenomena have been largely explored from a geophysical point of view, since frictional dynamics plays an important role in seismic faulting. Marone [3] has provided an excellent review of the extensive laboratory studies at high pressures that have been made in studies too numerous to cite here. One of the central achievements of these efforts has been the development of friction laws that incorporate memory effects and history dependence; frictional forces are not simply determined by the relative velocity of the shearing surfaces. Rather, localized contacts tend to strengthen with age and weaken as they are replaced during motion. These effects can be summarized by phenomenological friction laws depending on velocity and one or more additional state variables that characterize the macroscopic state of the contacting surfaces. Often a single state variable that can be regarded as a measure of average contact age is sufficient. An earlier review of these friction laws may be found in the book by Scholz [4]. Additional studies of granular friction at pressures relevant to seismic phenomena may be found in Refs. [5-9].

Sensitive studies of solid-on-solid friction at *low* pressures (without granular material) have also been conducted, especially by Heslot, Baumberger, Perrin, and their collaborators [10-12]. These studies have highlighted the fact that there is generally not an absolute distinction between static and dynamic friction. Surfaces subjected to a shear stress can exhibit slow evolution and deformation even when there is little evidence of macroscopic motion. In some cases this slow creep can apparently be thermally activated. As is now well known, the elastic properties of the apparatus influence the observed frictional behavior. It is also interesting to note the experiments of Johansen et al. [13], who studied the behavior of a metal-on-metal spring-block system and noted clearly *nonperiodic* stick-slip dynamics if the normal force is not too large. A much older study of solid-on-solid friction due to Tolstoi [14] demonstrated the importance of vertical motion in solid friction. For a general review on sliding friction see the recent book by Persson [15].

The nonlinear dynamics of block-spring systems under the influence of typical velocity-dependent friction laws has been recently discussed theoretically by Elmer [16], who showed that in addition to the usual stick-slip and steady motions, an oscillatory state without sticking is also possible. An earlier theoretical study of such a system using more complex rate and state dependent friction laws was given by Rice and Tse [17].

A considerable body of work beyond that already referenced is at least indirectly related to this investigation. Force distributions in granular matter under normal and/or shear stresses have been studied both experimentally and theoretically [18-20]. The forces supported by the particles are not homogeneous in space but are concentrated into "stress chains" that are quite important in sustaining frictional forces [2, 21]. Discrete particle simulations have been used to study friction both in the quasi static regime where particles are usually in contact, and in the rapid flow regime that is often modeled using kinetic theory methods [22]. The strong density dependence of friction in granular media has been studied experimentally by Horváth et al. [23]. Dilatancy has been shown to affect the onset of flows on an inclined bed [24]. Fluidization of a granular material excited by horizontal vibration has been studied experimentally [25] and may be related to the fluidization that occurs due to shear in the present investigation. A simple one-dimensional model of spring coupled particles interacting with a random substrate [26] reveals complex spatiotemporal dynamics. Another one-dimensional model, but involving a periodic potential, emphasizes the analysis of the relaxation behavior during the slip events [27].

It is interesting to consider the parallels between friction in granular layers and the study of friction between surfaces separated by a thin layer of lubricant, which has provided many surprises and insights. In that domain, sensitive experimental methods have demonstrated the profound effects of molecular discreteness. Changes in the state of organization of the lubricant layer, e.g. shear-induced melting or pressure-induced "solidification", are known to cause qualitative transitions in the frictional dynamics as the conditions are varied. As examples of these studies, we mention Refs. [28-32]. Phenomenological friction laws capable of describing some of these phenomena have been discussed recently [33].

**The present work:** Here we report an extensive study of sheared granular layers at low normal stresses, in part using methods similar to those of Refs. [11, 12]. A brief preliminary report on this study appeared previously [34]. This work is distinguished from earlier investigations involving granular materials in several ways. First, we are able to determine the temporal variations of the frictional force within the individual slip events lasting only tens of milliseconds. As we will show, the frictional force is multivalued, with the instantaneous force being less for decreasing than for increasing velocities. Second, we are able to detect microscopic particle rearrangement events optically, and to correlate these local events with the slow global creep that precedes (and follows) major slipping events. These "precursors" and "afterslip" events seem analogous in some respects to those that occur on some faults before and after major earthquakes [35-37]. An advantage of studying friction at low pressures is that particle geometry, but not plasticity, should play the dominant role. In addition, changes in the particle size distribution and shape do not occur.

Besides providing a more thorough treatment here, we go beyond our previous brief report by showing in some detail how the stick-slip dynamics depends on the shear rate and on the strength of the spring that connects the translating plate to the external drive mechanism. We also measure explicitly the dilation of the material which begins prior to slip and which is the major cause of the observed hysteresis in the friction force. This effect has been noted in "molecular" dynamics simulations of granular materials by Thompson and Grest, and Zhang and Campbell [22, 38]. Finally, our extensive work on smooth glass particles is supplemented by some studies of rough particles, though this aspect of the investigation will require further work before it can be considered complete, since several different aspects of the material properties can influence the results.

Following a discussion of experimental methods in Sec. II (including the measurement apparatus, samples, imaging, and methods of resolving the actual frictional force), we discuss the experimental results in Sec. III. The velocity dependence of the stick-slip behavior, the structure of the individual events, material dilation, microscopic precursors, and the behavior of rough particles are considered in turn. We summarize the observations and compare the results to those of other investigations in Sec. IV.

## II. Experimental Methods

### A. Apparatus

A schematic diagram of the experimental setup is shown in Fig. 1. Shear stress is imposed by translating a transparent cover plate over a uniform granular layer. The horizontal dimensions of the upper plate are generally 75 x 50 mm; the lower one is much larger. The edges of the plate are rounded off to avoid the possibility of it digging into the granular layer, but this precaution is found to have little or no effect on the measurements.

In order to control the stiffness of the driving system, the cover plate is pushed with a blue tempered steel leaf spring (thickness 0.005"-0.025", length <1", width 0.5", and spring constant $k$) connected to a translating stage. The stage is driven at a constant speed $V$ by a computer-controlled stepping motor through a high precision micrometer. The coupling between the spring and the cover plate is accomplished through a 1/16" diameter stainless-steel ball, which is glued to one end of the cover plate. This allows the plate to move horizontally and also vertically as the thickness of the granular layer may vary during the motion. The coupling method ensures that the external force does not exert a torque on the plate about its center-of-mass.

The elastic coupling through the spring allows relative motion of the top plate with respect to the translator. The relative motion in the horizontal direction is monitored by measuring the displacement $\delta x(t)$ of the spring from its rest position at the coupling point with an inductive position sensor, model EMD1050 from Electro Corporation as in Ref. [12]. The horizontal position of the top plate in the laboratory frame is then given by $x(t) = Vt - \delta x(t)$. The output from the sensor is a weakly nonlinear function of the distance between spring and the sensor head. In addition, because of the bending of the spring, it doesn't remain perfectly vertical with respect to the sensor head, and this leads to a possible additional source of systematic error. Both sources of systematic error are eliminated by using a calibration function to obtain the actual spring deformation $\delta x(t)$. This calibration function is accurately determined for each spring and is then applied to the measured data. We estimate that the precision of the resulting deformation data is about 0.1 $\mu$m. This is mostly determined by electronic noise.

The deformation of the spring is sufficiently small that the force exerted on the top plate by the spring is given accurately by $F = k\delta x$. We also monitor the vertical displacement of the cover plate with a separate inductive sensor.

The granular sandwich is mounted on a microscope to allow the observation of microscopic motion of granular particles through the transparent top plate. The particle motion can be captured on video tape for later quantitative analysis, or digitized directly. Synchronization allows events in the images to be correlated with structure in the force measurements. The apparatus and microscope are contained within a temperature-controlled box and maintained somewhat above room temperature (typically 37.5±0.2 °C) to reduce the extent of adsorbed water on the particles. The humidity during the experiments is typically 20±2 %. Without this control, we found unacceptable variations from run to run. However, it is likely that no experiment performed in air can completely eliminate the effects of adsorbed water.

Detailed studies reported here utilized the following ranges of the major parameters: imposed translation or driving velocity $V$ of the spring mount ($10^0$-$10^4$ $\mu$m/s); layer thickness (2 mm); spring constant $k$ ($10^2$-$10^4$ N/m); mass $M$ of the upper plate (10-50 g). Edge effects (due to the finite horizontal size of the layer) are believed to be negligible.

### B. Samples and surface treatment

We have explored samples consisting of (a) smooth spherical glass particles 70-110 $\mu$m in diameter (obtained from Jaygo, Inc.) and (b) clean art sand consisting of rough particles 100-600 $\mu$m in diameter (obtained from Estes Inc.). Images of these two samples are shown in Fig. 2. For our samples, cohesive forces are negligible. Samples are used only for a few runs to avoid the possible effects of wear, though we have not observed such effects. In this paper, we focus primarily on the behavior of the spherical particles, and briefly summarize the behavior of the rough particles. A further paper on the rough particles and other materials is anticipated.

Before each run the granular layer is distributed on the bottom plate so that the depth $h$ is constant over the whole layer (generally 2.00±0.05 mm for the spherical particles; 4.0 mm for the rough particles), and then the top plate is placed slowly upon it. Sometimes the upper plate sinks slightly into the granular layer as the plate translates. This can lead to a small heap at the leading edge, though the effect is minimized by avoiding torques due to the driving force and by keeping the mass of the plate small. In the most recent experiments, we bevel the leading edge of the glass plate to limit the accumulation of material there. We also checked that there is no perceptible difference between data acquired early in a run and that obtained late in a run. The mass of any displaced material is small compared to the mass of the top plate, and the resulting drag is also minimal because the material is unconfined.

The surfaces of the plates are treated in several ways to transmit the shear force to the granular layer without interfacial slipping. We roughen the bottom plate and either (a) glue a layer of particles to a glass top plate, or (b) rule a Plexiglass top plate with parallel grooves 2 mm apart, so that imaging measurements can be made in the interstices. Similar results are found in the two cases.

### C. Instantaneous velocity and frictional force measurements

The instantaneous horizontal velocity of the top plate can be obtained simply by differentiating its position $x(t) = Vt - \delta x(t)$. The time resolution depends on the sampling rate, which is typically 1000 Hz. We can also determine the velocity dependence of the instantaneous normalized frictional force $\mu(t) = F_f(t)/Mg$ exerted on the top plate by the granular layer. Here $F_f(t)$ is the actual frictional force, $M$ is the mass of the cover plate, and $g$ is the gravitational acceleration. This quantity is derived from the data by first obtaining the position $x(t)$ and the acceleration $\ddot{x}(t)$ of the top plate from the measured spring deflection $\delta x(t) = Vt - x(t)$. Some digital filtering of the differentiated data is required, so there is a tradeoff between signal-to-noise and time resolution. The acceleration of the top plate is determined by the difference between applied and frictional forces. Therefore, we use the force balance equation

$$M\ddot{x} = k\delta x - F_f \tag{1}$$

to obtain the *instantaneous* frictional force as a function of either time or the instantaneous sliding speed $v(t) = \dot{x}(t)$.

### D. Image analysis

To study the dynamics of particle motion in the upper few layers of material, we digitize microscopic images of the particles using a standard frame grabber and a shuttered CCD camera. The resolution of the image is 640 x 480 pixels, with image depth of 8 bits. We use this information mainly to search for local rearrangement events. These small motions can be detected by taking differences between images separated in time by a suitable interval, typically 1 s. To count (approximately) the number of particles that have moved, we first convert the image to a black/white (1 bit depth) image by using a threshold. We coarse grain the image over a size comparable to the mean particle diameter. The number of spots for which the difference image is non-zero is a measure of the number of particles that have been displaced between the two images.

## III. Experimental Results

### A. Low strain-rate stick-slip phenomena

We concentrate on the behavior of the material containing spherical particles, and begin with an example for which the stiffness $k$ is low, 134.7 N/m. For low driving velocities, the top plate alternately sticks and slips, as shown in Fig. 3. During the sticking periods, the top plate is at rest, and the force $F$ exerted on the top plate (proportional to the displacement $\delta x$ of the free end of the spring) increases linearly in time up to a maximum value $F_s$. When $F$ exceeds $F_s$, the granular layer no longer can sustain the external shear stress, and the top plate begins to slide. Visual observation through the top plate reveals that this onset of slipping involves fast irregular motion (which we term "fluidization") of granular particles near the top plate. During the sliding period, the top plate accelerates to catch up with the moving stage and $F$ decreases. At some lower value, the top plate sticks again. The slip duration $\tau_{slip}$ is roughly of the order of the inertial characteristic time $\tau_{in} = 2\pi\sqrt{M/k}$, and is substantially shorter than the sticking interval $\tau_{stick}$ in this regime. The granular flow during slip seems to be confined to the top few layers of granular particles, but we do not have a quantitative measure of the actual depth.

The position $x(t)$ of the top plate shown in Fig. 3(b) is a sequence of apparently equal steps with nearly instantaneous jumps between them. The slip velocity achieved during each slip event is almost the same (Fig. 3c). By magnifying $x(t)$, we find that there is in fact gradual motion or "creep" in the sticking intervals. This is shown in Fig. 4 for two sample slip events. We estimate that creep amounting to about 1% of the total slip displacement is typical. However, the amount of creep before a slip event is highly variable from one event to the next. The standard deviation of the total creep appears to be of the order of the mean value, though we do not have sufficiently long runs to make a quantitative measurement of the variability.

The observed stick-slip motion is almost periodic for wide range of parameter values. However, strongly non-periodic motion occurs at very low $V$ and large $k$, as shown in Fig. 5(a). When the stiffness is large, the mean period is short, and the displacement per cycle is small. Here the mean displacement is only about 2 $\mu$m, which is much less than the particle diameter. Under these conditions, the maximum static force is almost the same for each cycle, but the degree to which the stress is unloaded (and hence the time of re-sticking) varies from cycle to cycle. We find that by increasing $V$ by a factor of 5, the motion becomes approximately periodic as shown in Fig. 5(b) for the same stiffness.

Though the motion is irregular for high stiffness and low speeds, patterns do occur in Fig. 5(a). It appears that large slip events tend to be followed by smaller ones. This property is emphasized in Fig. 6, where a return map for the slip distance $\Delta x$ is shown. There is in fact a systematic anticorrelation of the slip distance between adjacent events, though there is a lot of scatter. This anticorrelation could be regarded as an approximate period doubling. It is interesting to note that the slope of a linear fit to the points in Fig. 6 is approximately -1, which is the threshold for period doubling. If we think of this state in the context of nonlinear dynamics, then by changing a parameter it should be possible to exhibit a bifurcation. However, the phenomenon has not been studied in detail.

We believe that the irregularity of these very small displacements is related to the huge number of microscopic arrangements that are possible in a disordered medium, and also possibly to the tendency to form localized chains of particles that sustain significant shear stress [1, 18]. Although the number of particles in a layer is large (about $10^5$ per layer), fluctuations are not eliminated. Even the apparently periodic motion of Fig. 5(b) or Fig. 3 is not perfectly periodic, as shown by the variability of the creep process (Fig. 4b).

### B. Velocity dependence of stick-slip oscillations

We return to the case of somewhat lower stiffness, and consider the effect of varying the driving velocity. We present results for several different values of $V$ in Fig. 7. The first example is similar to Fig. 3(a). Though the motion is nearly periodic, slight fluctuations in amplitude and period are visible in Fig. 7(a). For sufficiently low $V$, the mean period $T$ of the stick-slip motion is proportional to $V^{-1}$; this dependence of $T$ on $V$ is shown by the straight line portions of Fig. 8. The period also decreases with increasing $k$. The product $VT$ corresponds to the mean distance traveled by the stage during the mean period $T$, and hence it is equal to the mean slip amplitude $A$. Since $T \propto V^{-1}$ for low $V$, the mean slip amplitude $A$ should be independent of $V$, and this is also observed, as we show in Fig. 9.

#### 1. Inertial effects

At higher driving speeds, e.g. for $V$>500 μm/s (and $k$=1077 N/m), the slip duration is no longer negligible with respect to the period $T$. As $V$ is increased, the periodic motion becomes smoother, though not quite sinusoidal (Fig. 7b); the upper plate still sticks briefly once per cycle. The period $T$ declines more slowly with increasing $V$ and eventually saturates at a value $T_{min}$. By changing the weight $M$ of the top plate and the stiffness $k$ of the spring, $T_{min}$ is found to be proportional to $\sqrt{M/k}$ and somewhat larger than the characteristic inertial time $\tau_{in}=2\pi\sqrt{M/k}$. We can describe the deviation of $T_{min}$ from $\tau_{in}$ by introducing an effective mass $M_{eff}$>M. For example, $M_{eff}$ =1.7$M$ for $k$=1077 N/m and $M$ =1.09 x $10^{-2}$ kg. These facts indicate that inertial effects are prominent in this regime. The peak-to-peak spring deflection amplitude $\delta x_{p-p} \equiv A$ increases monotonically with $V$ in this regime, as shown in Fig. 9. (For low $V$, this quantity is close to the peak-to-peak displacement amplitude of the top plate.)

#### 2. Transition to continuous sliding motion

If $V$ is increased sufficiently, the periodic motion is replaced by steady sliding motion with small intrinsic fluctuations (Fig. 7c). This transition can be seen as a sharp decrease in the peak-to-peak deflection amplitude for the lower stiffness runs in Fig. 9. In the vicinity of the transition, noisy oscillations are visible, but their amplitude is strongly modulated, as shown in Fig. 10. Because of these fluctuations, we use the mean value of $\delta x_{p-p}$ when computing *A*.

We find that the transition threshold $V_c$ tends to decrease with *k*. Though the transition appears to be discontinuous for $k<5000$ N/m, it is more gradual when *k* is increased, and then there is no pronounced drop in *A* (Fig. 9).

### C. Time-dependent velocity measurements

We now consider the instantaneous velocity variations of the moving plate during individual slip events as shown in Fig. 11. Recall that this quantity is obtained by differentiating *x(t)*. It is remarkable that the behavior during slip for different *V* is nearly identical when *k* and *M* are fixed. Also note that the acceleration and deceleration portions of the pulse are different. The maximum slip speed is of the order of $F_s/\sqrt{kM}$ and decreases with increasing *k*. The slip duration $\tau_{slip}$ is about 40 ms for $k=135$ N/m and $M=10.9$ g; it is of the order of the inertial time defined previously.

Velocity pulses for two different driving speeds *V* are shown in Fig. 12. The larger one is in the inertial regime, and the two pulses are made to coincide at *t*=0. . Note that the falling side of the pulses approach a straight line, especially for the larger pulse. This linearity suggests that the frictional force from the granular layer is approximately constant as the upper plate decelerates, while the curvature of the pulses on the increasing side implies that the force is *not* constant as the upper plate accelerates. However, to be certain of these conclusions it is necessary to correct for the time variations of the spring force, as we do in the next section.

### D. Instantaneous frictional force measurements

We explained in Sec. II.C how the instantaneous frictional force from the granular layer can be determined. In Fig. 13, we show the normalized instantaneous frictional force $\mu(t)=F_f/Mg$, calculated from the deflection signal and plotted as a function of the instantaneous velocity $\dot{x}(t)$. During the sticking interval, $\mu$ increases linearly in time from A to B. When $\mu$ reaches the static threshold $\mu_s = F_s/Mg$, the top plate starts to slide. As can be seen from Fig. 13, the frictional force during acceleration is distinct from that during deceleration. During acceleration, $\mu$ decreases monotonically from $\mu_s$ with the increase in the slip velocity (from B to C in Fig. 13). During deceleration, $\mu$ is nearly constant for some time starting from C, and then drops quickly for small speeds to the resticking at A. This loop is almost identical for all slip events, independent of *V*.

For the inertia-dominated oscillations, the loop depends on driving speed. In this regime, creep is detectable for $\mu<\mu_s$ and a clear distinction between stick and slip becomes more difficult to make; see the loop ABC' in Fig. 13. The maximum slip speed increases with *V*, but the frictional force $\mu$ during rapid motion (near the maximum slip velocity, at C') is almost independent of *V*.

These observations imply that the time-dependent frictional force is determined not only by the instantaneous velocity, but *also* by the history of the motion, or by a state variable such as the degree of fluidization. One might *attempt* to explain the observations by introducing a delay or memory effect, as was done long ago by Rabinowicz [39] to describe hysteresis loops observed in solid-on-solid friction; see also Ref. [40]. In that approach, the instantaneous frictional force $F_f$ at time $t$ would be determined uniquely by the velocity at an *earlier* time ($t$-$\tau_o$) through a single-valued, monotonically declining function $F_o(v)$:

$$F_f = F_0(\dot{x}(t - \tau_0))$$
$$= F_0\left(V - \partial_t \delta x(t - \tau_0)\right) \quad .$$

The combination of a delay and a declining dependence on velocity would produce hysteresis if $F_f$ is plotted as a function of the velocity $\dot{x}(t)$. Note that if $V << \partial_t \delta x(t - \tau_o)$ during the motion, then $V$ could be neglected and the dynamics approximated as being independent of $V$ for part of the cycle, as we have observed. However, this approach *does not work*; a fixed delay *cannot* explain the observed rapid fall in $F_f$ at very small $v$. This observation implies that an additional state variable is probably necessary. We return to this point in the concluding section.

### E. Dilation during slip pulses

Simulations [22, 38] suggest that dilation and fluidization play a role in the stick-slip instability. Microscopic visualization in our experiment reveals that the granular particles in the upper few layers flow with the top plate during the slip period. Although dilation is known to occur for high speed continuous motion, it has not previously been detected for individual slip events. By using an additional vertical displacement sensor, we are able to detect small dilation pulses associated with each slip event. The time-dependent vertical displacement is shown in the stick-slip regime in Fig. 14. Unfortunately our sensitivity is less (due to instrumental noise) for the vertical displacement. However, we can estimate that the amplitude of the dilation is about 15 µm, substantially less than the mean particle diameter. (The pulses appear somewhat smaller in Fig. 14 because of filtering that was applied to the signal to improve the signal-to-noise ratio.) It should be noted that the mean depth is about 20 particle diameters, and that the dilation is probably largely near the upper surface. In Fig. 14 we also show the corresponding horizontal velocity during the pulses. The duration of the dilation pulse is somewhat broader than the velocity pulse, and it starts earlier. The peak of the dilation clearly occurs *before* the peak in the velocity. Further measurements indicate that the dilation peak coincides approximately with the maximum *acceleration*. This fact makes qualitative sense if the dilation determines the frictional force once slipping starts.

### F. Microscopic precursors

Optical imaging allows us to search for localized particle rearrangements between major slip events, as explained in Sec. II.D. To allow optical access, we use a ruled Plexiglass plate as discussed earlier. We find that local rearrangements of granular particles occur even during the sticking intervals. An image of a portion of the layer (about 10 grains or 1 mm in width) is shown in Fig. 15, along with a sample difference image showing local displacements in 1 s (circled). We find that the microscopic rearrangement rate, as indicated by average number $<n(t)>$ of microscopic displacement sites (in a 1 s interval and a 4 mm$^2$ area) increases sharply near the time of a major slip event. This time dependence is shown in Fig. 16(a). The data has been averaged over 416 slip events. Note that the event rate is asymmetric, with precursors being somewhat more likely than microscopic events at an equivalent time *after* a major slip occurs at *t/T=0*.

There is substantial variability in the number of microscopic rearrangements before (or after) a slip, just as there is variability in the macroscopic creep (Fig.4). This variability may be described by the probability distribution *P(N)* of the *total* number *N* of microscopic slip sites at *any* time before or after a single major slip event. This quantity, shown in Fig. 16(b), is approximately an exponential function with a decay time of 40 events, comparable to the mean value. The most likely explanation for the local re-arrangment events is the breaking of stress chains. Since the stress is highly inhomogeneous at low *V*, a few local events can allow measurable global motion.

### G. Results for rough particles

An extensive discussion of the shear friction for rough particles will be given in a subsequent paper, but we summarize the main effects here. The most important difference with respect to the spherical particles is that approximately periodic motion is never found; the dynamics is always irregular. This is shown in Fig. 17, where we display the temporal evolution of spring displacement signals for several different conditions. In Fig. 17(a), for example, one can see several small slips before each major one; after each small slip, the spring deflection (and therefore the frictional force) rises slightly higher than the previous peak. These small slips are probably due to the breaking of some of the stress chains. This statistical behavior is reminiscent of that of avalanches.

In Fig. 17(b), we show the effect of increasing the driving velocity. The motion is even more irregular, and a well defined maximum static stress does not exist. Some major slip events are preceded by a number of smaller ones, while other slip events are not. Finally, we show in Fig. 17(c) an example of the dynamics for a weaker spring constant, with a driving velocity similar to that of Fig.17(a). Again, the major slips are irregular, though their larger mean size makes the small slips (the minor glitches) less obvious.

In general, we find a broad range of event sizes (slip displacements) for the rough particles. However, it appears that there may typically be a gap between the large events and the small ones. Additional observations to be reported in a separate paper show that successive slip events leave the plate at a slightly different vertical height, so that the granular material is in a slightly different state (or mean density). Finally, the high velocity inertial "resonance" seen in Fig. 9 for small *k* is not apparent in the rough material. At present, the limited total length of translation available in our apparatus restricts our ability to gather adequate statistics on the stochastic behavior of the dynamics involving rough particles.

## IV.  Discussion and Conclusion

### A.  Summary of observations

We have reported sensitive and fast force measurements on sheared granular layers undergoing stick-slip motion, along with simultaneous optical imaging.  Most of the work reported here pertains to spherical glass particles with a +-20% size distribution; limited measurements have also been made for rough particles.  Though we measure only the deflection of the coupling spring, we are able to compute from the measured data the slider's position, instantaneous velocity and acceleration, as well as the instantaneous frictional force produced by (or within) the layer.  The time resolution of the measurements is high enough to resolve the detailed dynamics within individual slip events lasting about 40 ms.

The effects of varying the elastic coupling constant $k$ and the driving velocity $V$ are studied in detail.  Stick-slip motion (Figs. 3,5) occurs for low driving velocity .  Between the slip pulses, we find slight creep (Fig. 4) that is highly variable from one event to the next.  The total displacement associated with the creep is much less than that during the slips.  When the stiffness is large (and the driving velocity small), the stick-slip motion becomes irregular even for the smooth particles, and the mean displacement (about $2\mu$m) is substantially less than the mean particle diameter. The maximum static force (Fig. 5a) is almost the same from cycle to cycle, but the minimum frictional force fluctuates strongly.  We believe that these observations are worthy of theoretical attention, since one might have thought that the large number of particles per layer (about $10^5$) would be enough to average out such fluctuations.  The motion in this regime seems to involve both deterministic nonlinear phenomena (approximate period doubling) and stochastic aspects (variability).

At larger driving velocity $V$, the oscillations become smoother and are dominated by inertia (Fig. 7).  The period $T$ initially declines as $V^{-1}$ and then saturates (Fig. 8).  The amplitude $A$ of the relative motion (Fig. 9) grows with $V$ and then declines sharply at the onset of continuous sliding motion..  This transition at $V_c$ appears to be discontinuous for $k$ not too large, but may be continuous for large $k$.  We observe large fluctuations in the oscillation amplitude in the neighborhood of the transition (Fig. 10). It may be productive to think of the system as a noise-driven resonant system in this regime.   Perhaps this oscillation is related to that noted in the stability analysis of Elmer [16].

The time dependence of the instantaneous velocity of the upper plate (Fig. 11) and the frictional force exerted by the granular layer are determined within individual slipping events.  The frictional force is found to be a multi-valued function of the instantaneous velocity during slip, with pronounced hysteresis, and a sudden drop just prior to resticking (Fig. 13).

Measurements of vertical displacement reveal a very small dilation of the material (typically 15 $\mu m$) associated with each slip event (Fig. 14); the kinetic frictional force appears to reach its minimum value when the dilation is greatest.   We believe that improved measurements of vertical dilation will be very helpful in understanding the force data.

We used optical imaging to reveal localized microscopic rearrangements that occur before and after each macroscopic slip. Their number is highly variable, and has a roughly exponential distribution (Fig. 16(b)). It seems probable that the accumulation of these local displacements is the microscopic manifestation of macroscopic creep. It may be that the slip events result from the spatial growth of the local rearrangements, but we are unable to test this hypothesis without a much faster imaging system.

Measurements on rough particles show significant differences from those made on spherical particles. Most notably, the stick-slip motion at low $V$ is always nonperiodic; both small and large slip events occur. The force hysteresis loops are also different on each cycle.

### B. Comparison to solid-on-solid friction, lubrication, simulations, and geophysical studies

In many respects, the results of this investigation parallel those previously reported for solid-on-solid friction on rough surfaces [10-12]. The similarities include: the occurrence of stick-slip motion at low driving speed; a transition to continuous sliding motion at a critical speed that depends on stiffness; transitional behavior that switches from being discontinuous at low stiffness but may be continuous at higher stiffness; and observable creep before major slip events. In vesry recent work to be reported elsewhere, we find a dependence of the maximum static force on the sticking time, i.e., a strengthening of microscopic contacts with time. This was attributed to thermal relaxation in the solid-on-solid friction work. Thermal effects related to plastic deformation may be much weaker here, while geometrical effects are probably larger.

Our results on stick-slip dynamics in granular materials also resemble those reported for delicate studies of lubricated films between smooth mica surfaces (for example, those of Ref. [30]. Those experiments have been interpreted in terms of a slip-induced melting of the lubricant, and resticking due to a time- and velocity-dependent re-freezing transition. Perhaps one can think of the shear-induced dilation of the granular material observed in our experiments as being analogous to the slip-induced melting of a thin lubricant layer. In both cases, geometrical constraints related to particle (or molecular) size are important.

A clear phase boundary between solid and "liquid" phases was seen in the two-dimensional numerical simulations of sheared granular layers in Ref. [22]. In fact, the conditions of our experiments are comparable to those of the simulation, except that the mass of the upper plate leads to a somewhat larger applied normal stress in the experiments, and the particles are not identical.

It may be possible to define a state variable $\theta$ that is related to the degree of fluidization. Then one could describe the frictional force in terms of $\theta$ and the instantaneous slider velocity $\dot{x}(t)$, together with the evolution equation for $\theta$. With improvements in our measurements of the vertical dilation, it might be possible to carry out this program, and hence to relate our work to the abundant literature on rock friction and sheared granular gouge [3], where the utility of rate- and state- dependent friction models has been amply demonstrated.

However, there are also clearly some significant differences. For example, the high normal forces used in most laboratory geophysical measurements cause the properties of the granular material (including the size distribution) to evolve in time, an effect that is nearly absent here.

It is provocative to note that some earthquakes show peaks of activity near the times of major slip events, as do some models of single strike-slip faults [37]. Creep and afterslip are also well known in geophysics; the latter can account for a significant fraction of the total energy release [36]. Though the conditions of our experiments are far from those relevant in geophysics, they do provide fundamental information about the dynamics of sudden slip events involving granular materials under stress, and about localized precursors.

### D. Future work

We have varied many aspects of the experiments, including the translation velocity, spring constant, mass of the upper plate, and the mean size and roughness of the particles. We have taken care to minimize extraneous factors that could affect the results, including humidity variations and edge effects. We have taken care that the cover plate is sufficiently rough to pin the first layer. The qualitative effects reported here are robust.

On the other hand, the investigation leaves many questions open for the future. Optical studies with high time resolution may give insight into the nucleation of slip events. The velocity dependence of the frictional force should also be studied at high stiffness where the motion is continuous. (Our measurement sensitivity declines at high stiffness so a different apparatus is needed for this purpose.) The velocity dependence of the "static" friction coefficient is apparently rather small, but needs to be measured more accurately. The work should be extended to higher normal forces, and the statistical properties of the fluctuations in the regime of continuous motion deserve study. Much more can be usefully done to characterize friction in materials of different types. For example, we have not yet studied the effect either of an extremely wide distribution of particle sizes, or of strictly uniform particles. Also, we do not know whether the creep observed here is related to that observed in solid-solid friction; we suspect it is quite different because of the role of stress chains in granular material.

On the theoretical side, models of the hysteretic frictional force and its relation to the very small dilation (less than one particle diameter) are needed. The persistence of stochastic effects in the macroscopic measurements even for smooth materials deserves consideration. The statistical properties of the fluctuating slip dynamics of rough materials can perhaps be modeled. It may be possible to explore some of these phenomena numerically in three dimensions. Impressive codes for studying sheared granular materials have been developed [41].

## Acknowledgments

This work was supported in part by the National Science Foundation under grants DMR-9319973 and DMR-9704301 and by a Grant-in-Aid for Scientific Research from the Ministry of Education, Science, Sports, and Culture of Japan, No. 09740317. J.G. appreciates the support of a Belkin Visiting Professorship at the Weizmann Institute of Science, where the writing of this paper was completed. We thank J.M. Carlson, G.S. Grest, and P. Molnar for helpful discussions and suggestions.

## References and Notes

\*  Electronic address: nasuno@ele.kyutech.ac.jp.

\#  Electronic address: akudroll@clarku.edu. Current address: Department of Physics, Clark University, Worcester MA 01610, USA.

\+  Electronic address: jgollub@haverford.edu.

# Figure Captions

**Fig. 1.** Schematic diagram of the apparatus, showing the granular layer GL and a transparent cover plate CP pushed by a leaf spring SP connected to a translating stage TS. An inductive sensor PS detects the deflection of the leaf spring. The microscope objective MO is also shown.

**Fig. 2.** Samples of (a) spherical glass particles 70-110 $\mu$m in diameter, and (b) clean sieved sand, with non-spherical rough particles 100-600 $\mu$m in diameter.

**Fig. 3.** (a) Spring deflection $\delta x(t)$ as a function of time for periodic stick-slip motion. (Spring constant $k$=134.7 N/m; mass $M$=10.90 g; $V$=113.33 $\mu$m/s.) (b) Position $x(t)$ and (c) instantaneous velocity of the top plate. Pulses corresponding to the slip events.

**Fig. 4.** Magnification showing displacement creep (of variable amount) before several major slip events. ( $k$=134.7 N/m; mass $M$=10.90 g; $V$=11.33 $\mu$m/s.) The origin of the slider position is selected to be far from the slip events.

**Fig. 5.** (a) Irregular stick-slip motion at high stiffness ($V$=11.33 $\mu$m/s, $k$=3,636N/m, $M$=10.90 g). (b) A nearly periodic case from the same run for $V$=56.67 $\mu$m/s.

**Fig. 6.** Return map for the slip distance $\Delta x$ for successive events obtained from Fig. 5(a). Though there is substantial scatter, large slips tend to be followed by small ones.

**Fig. 7.** (a) Stick-slip motion at $V$=5.67 $\mu$m/s. (b) Inertia-dominated oscillation at $V$=5.67 mm/s. (c) Steady sliding motion with fluctuations, at $V$=11.33 mm/s. ($k$= 1077 N/m; $M$= 10.90 g.)

**Fig. 8.** Mean period $T$ of the stick-slip motion as a function of driving velocity $V$ for different values of the spring constants $k$. At low $V$, $T$ declines as $V^{-1}$ ($M$=10.90 g).

**Fig. 9.** Mean peak-to-peak deflection amplitude $A$ as a function of the driving velocity $V$ for different values of the spring constant $k$. The transition to steady sliding motion is discontinuous for small $k$.

**Fig. 10.** Strongly modulated oscillations in the vicinity of the transition to steady sliding motion ($k$=1088 N/m; $M$= 10.90 g; $V$=5893 $\mu$m/s).

**Fig. 11.** Instantaneous velocity $v(t)$ of the cover plate during slippage for various driving speeds in the stick-slip regime ($k$=135 N/m; $M$=10.90 g). The time origins of the pulses are forced to agree at the end of each event.

**Fig. 12.** Comparison the shapes of two velocity pulses for different driving speeds. The larger one is in the inertial regime. ($k$=134.7 N/m; $M$=10.90 g.)

**Fig. 13.** The normalized instantaneous frictional force $\mu(t) = F_f/Mg$ as a function of velocity $v(t)$. (This should not be confused with the external applied force.) Solid

circles: stick-slip regime ($V$=113 µm/s); Open circles: inertia dominated regime ($V$=5.667 mm/s).

**Fig. 14.**  Vertical displacement (upper curve) and horizontal velocity (lower curve) of the top plate as a function of time. ($k$=135 N/m; $M$=11.33 g; $V$=226.7 µm/s).  Note that the maximum vertical displacement precedes the maximum horizontal velocity.

**Fig. 15.**  (a) Image of a portion of a granular layer (about 1 mm across), and (b) difference image ($\Delta t$ =1 s) showing localized particle rearrangements (circled) between major slip events.

**Fig. 16.**  (a) Average number of displacement sites (per second in a 4 mm$^2$ area) as a function of time before or after a major slip event, in units of the period $T$ of the stick-slip cycle. Precursory and post-slip events are prominent.  (b) Probability distribution of the total number of displacement sites observed at any time before or after a single slip event.

**Fig. 17.**  Spring deflections $\delta x(t)$ as a function of time for rough particles (Estes art sand). (a) Spring constant $k$ = 1568 N/m; mass $M$=13.47 x 10$^{-3}$ kg; velocity $V$= 5.636 µm/s. (b) Larger velocity $V$ =112.7 µm/s. (c) Lower spring constant $k$ = 220.9 N/m ($M$=11.27 x 10$^{-3}$ kg; $V$=11.27 µm/s).  The motion is much more irregular than that of the glass spheres.